# Electronic band structure, Fermi surface, and elastic properties of new 4.2K superconductor SrPtAs from first-principles calculations

I.R. Shein, * A.L. Ivanovskii

*Institute of Solid State Chemistry, Ural Branch of the Russian Academy of Sciences, 620990 Ekaterinburg, Russia*

A B S T R A C T

The hexagonal phase SrPtAs (s.g. *P6/mmm*; #194) with a honeycomb lattice structure very recently was declared as a new low-temperature ($T_C$ ~ 4.2K) superconductor. Here by means of first-principles calculations the optimized structural parameters, electronic bands, Fermi surface, total and partial densities of states, inter-atomic bonding picture, independent elastic constants, bulk and shear moduli for SrPtAs were obtained for the first time and analyzed in comparison with the related layered superconductor $SrPt_2As_2$.

* Corresponding author. Tel.: +7-3433745331; Fax: +7-343-3744495, *Russia*
E-mail address: shein@ihim.uran.ru (I.R. Shein).

***Introduction.*** Since the discovery of superconductivity in close proximity to magnetism in Fe-As layered materials, the family of transition-metal pnictides has attracted tremendous attention, see reviews [1-10]. In this context the situation for the system Sr-Pt-As seems very interesting. Indeed, in 2010 the superconductivity ($T_c \sim$ 5.2K) for tetragonal 122-type phase SrPt$_2$As$_2$ was discovered [11], and very recently (2011) the same authors have reported the hexagonal phase SrPtAs as a new low-temperature ($T_C \sim$ 4.2K) superconductor, which is the first pnictide superconductor with a honeycomb lattice [12].

In this Communication we present the first-principles calculations which are performed to understand the features of the electronic and elastic properties for the newest superconductor SrPtAs. In result the optimized structural parameters, electronic bands, Fermi surface, total and partial densities of states, inter-atomic bonding picture, independent elastic constants, bulk and shear moduli for SrPtAs were obtained for the first time and analyzed in comparison with the same data for the related layered superconductor SrPt$_2$As$_2$ [13].

***Computational aspects.*** Our band-structure calculations were carried out by means of the full-potential method with mixed basis APW+lo (LAPW) implemented in the WIEN2k suite of programs [14]. The generalized gradient correction (GGA) to exchange-correlation potential in the PBE form [15] was used. The plane-wave expansion was taken to $R_{MT} \times K_{MAX}$ equal to 8, and the $k$ sampling with 16×16×6 $k$-points in the Brillouin zone was used. The calculations were performed with full-lattice optimization. The self-consistent calculations were considered to be converged when the difference in the total energy of the crystal did not exceed 0.1 mRy and the difference in the total electronic charge did not exceed 0.001 $e$ as calculated at consecutive steps. The hybridization effects were analyzed using the densities of states (DOSs), which were obtained by the modified tetrahedron method [16]. Furthermore, for the calculations of the elastic parameters of the considered SrPtAs phase we employed the Vienna *ab initio* simulation package (VASP) in projector augmented waves (PAW) formalism [17,18]. Exchange and correlation were described by a nonlocal correction for LDA in the form of GGA [15]. The kinetic energy cutoff of 500 eV and k-mesh of 16×16×8 were used. The geometry optimization was performed with the force cutoff of 2 meV/Å.

***Structural properties.*** SrPtAs crystallizes in a hexagonal KZnAs-type structure, space group *P6/mmm*; #194 [19]. The atomic coordinates are Sr: 2*a* (0;0;0), Pt: 2*c* (⅓;⅔;¼), and As: 2*d* (⅔;⅓;¼). The structure of ThCr$_2$Si$_2$ can be schematically described as a sequence of two honeycomb sheets, where one of them is formed by Sr atoms, whereas the other (PtAs) plane sheet is formed by hexagons Pt$_3$As$_3$, see Fig. 1.

The calculations of total energy (E$_{tot}$) *versus* cell volume were carried out to determine the equilibrium structural parameters for the considered SrPtAs. The calculated values (*a* = 4.2976 Å and *c* = 9.0884 Å) are in reasonable agreement with the available experiments ($a$ = 4.246 Å and $c$ = 8.967 Å [19]; $a$ = 4.244 Å and $c$ = 8.989 Å [12]). Some divergences are related to the well-known overestimation of the lattice parameters within LDA-GGA based calculation methods. Note also that for the SrPtAs, the Pt-As bond lengths (in hexagons Pt$_3$As$_3$, ~ 2.48 Å) are smaller, than the Pt-As bond lengths (~ 2.60 Å) inside [As$_2$Pt$_2$] blocks, see [13].

***Electronic properties and Fermi surface.*** The calculated band structure and electronic densities of states (DOS) for the considered SrPtAs phase are shown in Figs. 2 and 3, respectively. We see that the As 4$p$ states occur between -5.8 eV and -3.2 eV with respect to the Fermi energy (E$_F$ = 0 eV), forming two DOSs peaks A and B, where peak A is mainly of 4$p_{x+y}$, whereas peak B is mainly of 4$p_z$ character. The bands between -3.2 eV and -2.0 eV (peak C) are mainly of the Pt 5$d$ character. The near-Fermi bands are of mixed of Pt 5$d$ + As 4$p$ type. The contributions from the valence states of Sr to the occupied bands are quite small.

The near-Fermi bands demonstrate (Fig. 2) a "mixed" character: simultaneously with quasi-flat bands along L-M and M-K, a series of high-dispersive Pt 5$d$ - like bands intersects the Fermi level between A and L, and H and A points. These features yield an multi-sheet Fermi surface of SrPtAs (Fig. 4) which consists of a set of hole and electronic sheets. The majority of them are hole-like, namely, concentric cylinders along the Γ-A direction, as well as the quasi-two-dimensional-type sheets at the corners of the Brillouin zone. Only very small closed electronic-like pocket is centered at K. Let us note that despite a set of differences, the Fermi surface for SrPt$_2$As$_2$ adopts also a quasi-two-dimensional type [13].

In order to clarify the near-Fermi electronic states in more detail, we give in Table 1 the values of total density of states at the Fermi level, N(E$_F$), and the atomic- and orbital- decomposed density N$^l$(E$_F$). Let us note that we have also examined the influence of relativistic effects (the spin-orbital interactions (SOC) within FLAPW) on the electronic properties of SrPtAs. It was found that SOC mainly results in energy shift and splitting of core and semi-core Pt states, which lay deeply under Fermi's level, whereas the common picture of valence bands, as well as the values of N(E$_F$), change quite little, see Table 1, where the estimations of the Sommerfeld constant ($\gamma$) and the Pauli paramagnetic susceptibility ($\chi$) for SrPtAs under the assumption of the free electron model are presented.

It is seen that for SrPtAs the main contribution to N(E$_F$) comes from the Pt 5$d$ states, with some additions of the As 4$p$ states. The similar situation was found also for the related phase SrPt$_2$As$_2$ [13].

According to the experimental data [11,12] the critical temperature decreases as going from SrPt$_2$As$_2$ ($T_C$ ~ 5.2K) to SrPtAs ($T_C$ ~ 4.2K), and this correlates with decreasing of N(E$_F$), see Table 1. Therefore in the framework of the BCS theory the observed decreasing of the $T_C$ in the sequence: SrPt$_2$As$_2$ → SrPtAs may be attributed to the changes of N(E$_F$).

***Inter-atomic bonding***. Using the usual oxidation numbers of atoms, the ionic formula of examined phase is: $Sr^{2+}Pt^{1+}As^{3-}$. Thus, the ionic bonding should be assumed between the adjacent $Sr^{2+}/(PtAs)^{2-}$ sheets as well as between the ions with opposite charges $Pt^{1+}/As^{3-}$ inside (PtAs) sheets. But our results show that the bonding picture for SrPtAs is more complicated and can be characterized as a high-anisotropic mixture of ionic and covalent contributions, see the shape of the isosurface of valence charge density, Fig. 1. Indeed, the covalent Pt-As bonds inside (PtAs) sheets (owing to hybridization of the Pt $5d$ - As $4p$ states, see site-projected DOSs, see Fig. 3) are well visible on the charge density map, Fig. 1.

***Elastic properties.*** The independent elastic constants ($C_{ij}$) for SrPtAs were evaluated by calculating the stress tensors on different deformations applied to the equilibrium lattice of the hexagonal unit cell, whereupon the dependence between the resulting energy change and the deformation was determined. All $C_{ij}$ are positive ($C_{11}$ = 148 GPa; $C_{12}$ = 95 GPa; $C_{13}$ = 36 GPa; $C_{33}$ = 85 GPa; $C_{44}$ = 27 GPa; $C_{66}$ = 26 GPa) and satisfy the well-known Born's criteria for mechanically stable hexagonal crystals: $C_{44}$ >0, $C_{11}$ > |$C_{12}$|, and ($C_{11}$ + $C_{12}$)$C_{33}$ > 2$C_{13}^2$. From the calculated constants $C_{ij}$, the elastic moduli for SrPtAs were evaluated, namely, the bulk modulus ($B$), compressibility ($β$) and shear modulus ($G$) - in three main approximations: Voigt (V) and Reuss (R) and Voigt-Reuss-Hill (VRH), see details in [13].

The calculated parameters are presented in Table 2 and allow us to make the following conclusions:

(i). the bulk modulus for SrPtAs is quite small (~ 45 GPa), and therefore this phase should be classified as a soft material with high compressibility ($β$ ~ 0.023 GPa$^{-1}$);

(ii). For SrPtAs it was found that $B > G$; this implies that the parameter limiting the mechanical stability of this phase (as well as for SrPt$_2$As$_2$ [13]) is the shear modulus $G$, which represents the resistance to shear deformation against external forces;

(iii). The widely used malleability parameter of materials is the Pugh's criterion ($G/B$ ratio) [20]. As is known, if $G/B$ < 0.5, a material behaves in a ductile manner, and *vice versa*, if $G/B$ > 0.5, a material demonstrates brittleness. In our case, according to this indicator, the phase SrPtAs lays on the border of brittle/ductile behavior - unlike SrPt$_2$As$_2$ which will behave in a ductile manner [13].

***Conclusions.*** In summary, by means of the first principles calculations, we have studied for the first time the structural, electronic and elastic properties of the hexagonal SrPtAs with a honeycomb lattice structure, which was reported very recently as new low-temperature superconductor.

The main conclusion is that this material should be characterized as quasi-two-dimensional ionic metal, which consists from metallic-like (PtAs) sheets alternating with Sr atomic sheets coupled by ionic interactions. The near-Fermi valence bands are derived from the Pt 5*d* states with an admixture of the As 4*p* states. The Fermi surface of SrPtAs is formed by a set of quasi-two-dimensional hole-like sheets parallel to the $k_z$ direction and by a single very small closed electronic-like pocket.

We also predicted that SrPtAs is a mechanically stable and soft material with high compressibility, where the parameter limiting the mechanical stability of this phase is the shear modulus *G*.

Finally, it is worthy to mention that the presented data (together with the earlier results [13]) allow us to speculate that the lowering the critical temperature in the sequence of related phases $SrPt_2As_2 \rightarrow SrPtAs$ may be attributed to the decreasing of $N(E_F)$.

**Acknowledgments.** This work was supported by the Russian Foundation for Basic Research, Grants No. RFBR- 09-03-00946 and No. RFBR- 10-03-96008.

**Table 1.** Total (in states/eV·f.u.) and partial (in states/eV·atom) densities of states at the Fermi level N($E_F$), electronic heat capacity $\gamma$ (in mJ·K$^{-2}$·mol$^{-1}$), and molar Pauli paramagnetic susceptibility $\chi$ (in $10^{-4}$ emu·mol$^{-1}$) for SrPtAs.

| parameters | I | II * |
|---|---|---|
| N($E_F$) total | 2.12 | 2.07 (2.55) |
| N($E_F$) Pt 5$d$ | 0.71 | 0.70 |
| N($E_F$) Pt 5$d_{z^2}$ | 0.03 | 0.03 |
| N($E_F$) Pt 5$d_{x^2+y^2+xy}$ | 0.21 | 0.20 |
| N($E_F$) Pt 5$d_{xz+yz}$ | 0.49 | 0.47 |
| N($E_F$) As 4$p$ | 0.32 | 0.30 |
| N($E_F$) As 4$p_z$ | 0.22 | 0.22 |
| N($E_F$) As 4$p_{x+y}$ | 0.08 | 0.08 |
| $\gamma$ | 5.00 | 4.88 (6.01) |
| $\chi$ | 0.68 | 0.67 (0.82) |

* as obtained within SOC (I) and without SOC (II)
** the data for related phase SrPt$_2$As$_2$ are given in parentheses [13].

**Table 2.** Calculated bulk modulus (*B*, in GPa), compressibility (*β*, in GPa$^{-1}$), shear modulus (*G*, in GPa), and Pugh's indicator (*G/B*) for SrPtAs in comparison with the same parameters for SrPt$_2$As$_2$ [13].

| phase/ parameter | SrPtAs | SrPt$_2$As$_2$ |
|---|---|---|
| $B_{V,R,VRH}$* | 79/10/44.5 | 101/99/100 |
| $\beta$ | 0.023 | 0.010 |
| $G_{V,R,RVH}$* | 30/15/22.5 | 27/25/26 |
| G/B | 0.51 | 0.26 |

* *B(G)*$_{V,R,RVH}$ - as calculated within Voigt (V)/Reuss (R)/Voigt-Reuss-Hill (VRH) approximations, as obtained within VASP

**FIGURES**

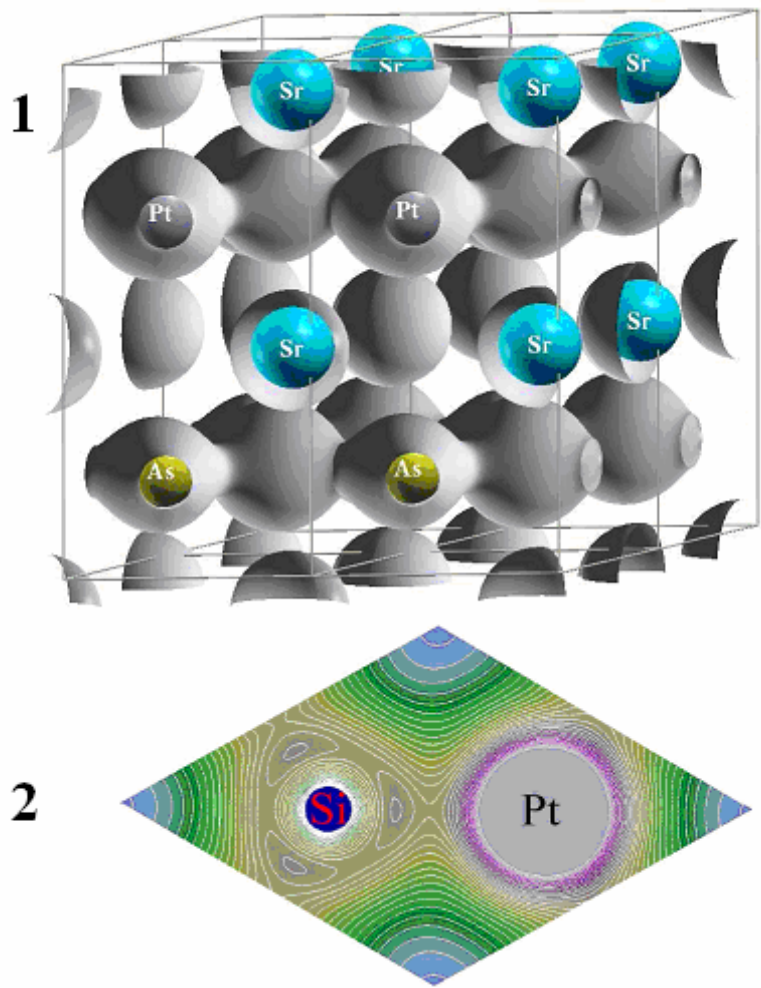

**Figure 1.** 1 - The iso-surface of valence charge density for SrPtAs, and 2 - charge density map illustrating the formation of directional Pt-As covalent bonds inside hexagons $Pt_3As_3$.

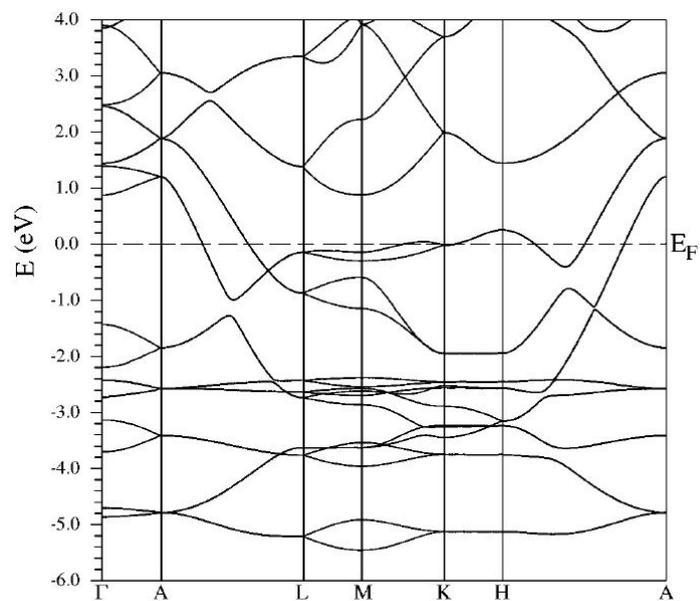

**Figure 2.** Electronic band structure of SrPtAs.

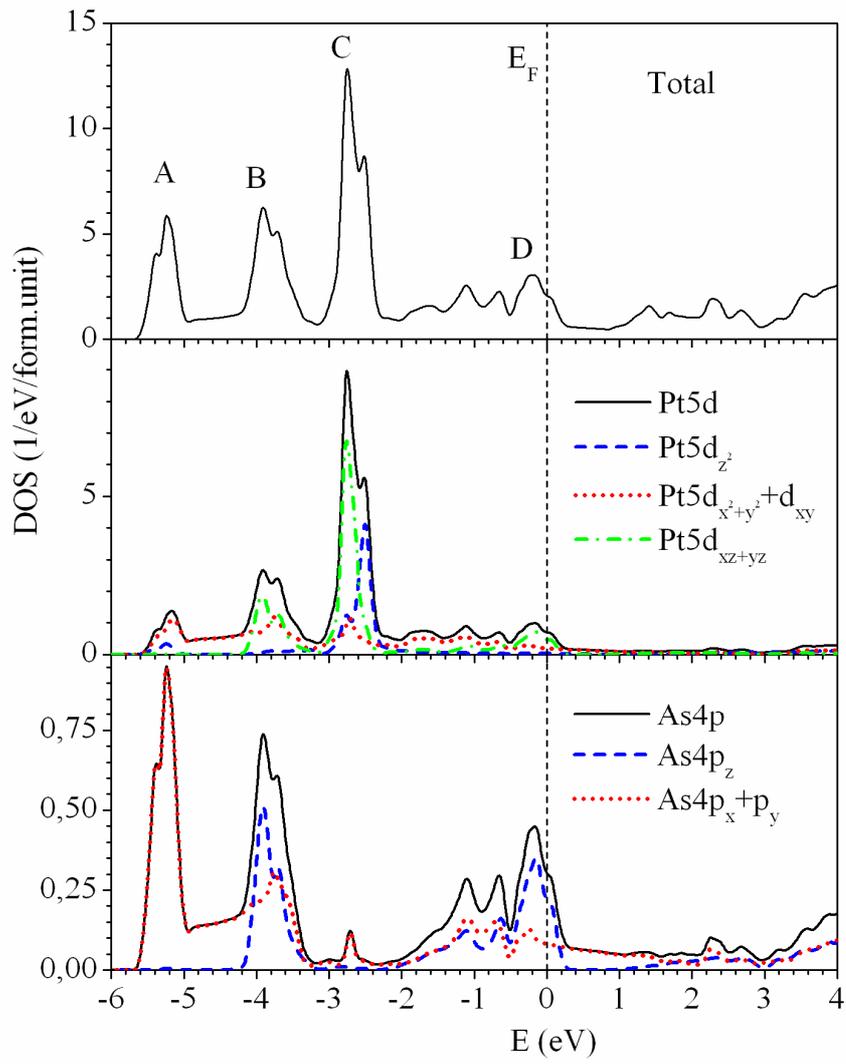

**Figure 3.** Total and partial densities of SrPtAs.

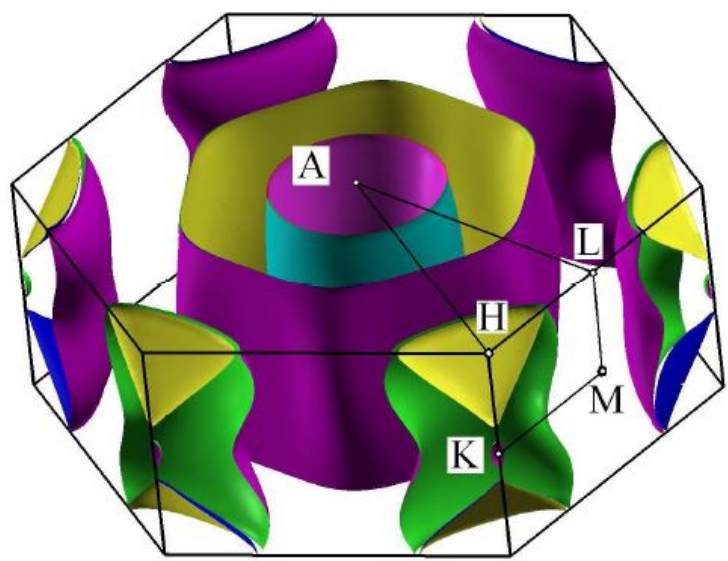

**Figure 4.** Fermi surface of SrPtAs.

8